\documentclass{article}

\usepackage{epsf}

\begin{document}

\begin{flushright}
hep-ph/0201119\\
\end{flushright}

\vskip\baselineskip

\begin{center}

{\large\bf SPIN AND OTHER ASPECTS OF \\[0.4em] GENERALIZED PARTON
DISTRIBUTIONS}\,\footnote{Talk given at Pacific Spin 2001, Beijing,
China, 8--13 Oct.~2001.  To appear in the Proceedings
(Int.\ J.\ Mod.\ Phys.\ A)} \\
\vskip 3\baselineskip

M.\ Diehl\,\footnote{Email: mdiehl@physik.rwth-aachen.de}
\\[0.5em]
{\small {\it Institut f\"ur Theoretische Physik E, RWTH Aachen, 52056
Aachen, Germany}}\\
\end{center}

\vskip\baselineskip

\begin{abstract}
I discuss how generalized parton distribution probe various aspects of
QCD bound states.  Topics include the interplay between transverse and
longitudinal structure, quantum mechanical interference, orbital
angular momentum, helicity flip, and higher-spin targets.
\end{abstract}

\section{Before talking about spin}

My assignment is to speak about spin aspects of generalized parton
distributions.  Before doing so I would however like to point out some
of the information they contain even if the spin degrees of freedom
are all averaged over.

\begin{figure}[htbp]
\begin{center}
\leavevmode
\epsfxsize=0.8\textwidth
\epsfbox{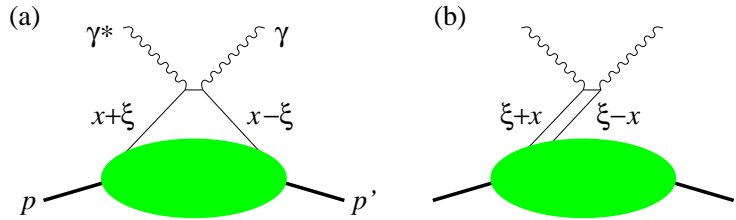}
\end{center}
\caption{\label{fig:handbag} One of the leading diagrams for virtual
Compton scattering in the limit of infinite photon virtuality $Q^2$ at
fixed Bjorken variable $x_B$ and invariant momentum transfer $t$.  A
second diagram is obtained by interchanging the photon vertices.  The
shaded blob represents a generalized quark distribution.  Cases (a)
and (b) correspond to different regions of the loop integration over
$x$.}
\end{figure}

The usual parton densities we are accustomed with are expectation
values of quark or gluon operators for a hadron state, and generalized
parton distributions
\cite{Muller:1994fv,Ji:1997ek,Rad:1997ki,Blum:1999sc} are their
natural extension to non-forward kinematics.  This is readily seen
from the diagrams for Compton scattering in the deeply virtual limit
(Fig.~\ref{fig:handbag}).  Since an off-shell photon turns into an
on-shell one, the two proton states in the hadronic matrix element
must have different momenta.  A useful set of variables describing the
kinematics are $x$ and $\xi$, parameterizing longitudinal momentum
fractions in a frame where both protons move fast, and the Mandelstam
invariant $t=(p'-p)^2$, which one can trade for the transverse
momentum transfer $(p'-p)_\perp$.  This tells us immediately that we
are in a position to probe the transverse and the longitudinal
structure of a fast moving proton in a correlated manner.

Setting $\xi$ and $t$ to zero one recovers the usual parton
distributions, which have the intuitive interpretation of a density in
longitudinal parton momentum.  Burkardt \cite{Burkardt:2000za} has
shown that for $\xi=0$ we retain a density interpretation if we
perform a Fourier transformation from $(p'-p)_\perp$ to impact
parameter, which measures the transverse distance of the struck quark
from the proton's center.  Generalized parton distributions at $\xi=0$
thus describe the transverse distribution of partons with a given
longitudinal momentum fraction $x$.  Going to nonzero $\xi$ we no
longer have this density interpretation.  I see this as a blessing
rather than a curse: after all QCD is a quantum theory, and it may
well be that to understand its bound states we will have to go beyond
classical probabilities.  In any case, generalized parton
distributions probe the transverse and longitudinal structure of the
proton in a way that has recently been described in the language of
photography \cite{Ralston:2001xs}.

The meaning of generalized parton distributions at nonzero $\xi$ can
be elucidated in terms of light-cone wave functions
\cite{Brodsky:2000xy}.  Barring possible subtleties of the 
light-cone gauge \cite{Brodsky:2001ue}, one can write the usual parton
densities as squared wave functions, summed over all configurations
containing a parton with given momentum fraction $x$.  Going to
$\xi\neq 0$ we get the product of wave functions corresponding to
different momentum fractions of the specified parton.  We thus go from
classical probabilities to quantum mechanical correlations, or
interference terms.  Nonzero $\xi$ further opens up a kinematical
regime where the distributions describe not the emission and
reabsorption of one parton (Fig.~\ref{fig:handbag}a) but rather the
emission of two partons (Fig.~\ref{fig:handbag}b).  One then probes
$q\bar{q}$ or gluon pairs in the target wave function.  Sum rules tell
us that the two regimes are intimately connected: both contribute to
the integral of the distributions over $x$ at given $\xi$ and $t$,
which gives the elastic proton form factors.

These sum rules remind us that quantities like the Dirac form factor
$F_1(t)$ also contain information about the proton size.  From the
above discussion it follows that they give the transverse distribution
of quarks, now averaged over all longitudinal momenta.  This is the
opposite reduction of information than the one we encountered in the
usual parton densities.  Generalized densities combine and interrelate
these two partial descriptions.  Note also that there are not many
pointlike probes providing local currents for a direct measurement of
form factors in the nucleon.  In particular there is none for gluons
to tell us about their transverse location compared to quarks.  Such
information may be gained in suitable diffractive processes where
generalized gluon distributions are enhanced over the quark ones.

We see that, even before talking about spin, we have access to a much
more detailed image of a hadron than the essentially one-dimensional
projection of usual parton densities and the information from form
factors.

\section{The two roles of spin}

Let us now investigate the spin degrees of freedom in deeply virtual
Compton scattering and in exclusive meson production $\gamma^* p\to M
p$, or in their counterparts $\gamma p\to \gamma^* p$ and $M p\to
\gamma^* p$ with timelike photons \cite{Berger:2001xd}. It turns out
that polarization for the photon and meson plays a quite different
role than on the target side.

\subsection{The photon side}

To access generalized parton distributions one needs a large
virtuality $Q^2$ of the virtual photon, and polarization offers a
valuable tool to asses whether a given $Q^2$ is indeed large enough.
This is because in the large limit $Q^2$ selection rules emerge for
the helicities on the photon side.  They follow from two ingredients.
First, one can neglect $t$ when evaluating the hard partonic
subprocess, where effects of nonzero $t$ are suppressed by a power of
$\sqrt{-t} /Q$.  The hard scattering thus becomes collinear.  The
second ingredient is that the chirality of light quarks is conserved
in the hard scattering, up to corrections in the tiny current quark
masses.  The axial anomaly of QCD does not invalidate this argument:
chirality violation can occur in quantities dominated by soft physics
but not in a hard scattering kernel \cite{Collins:1999un}.

Taking the Compton diagram in Fig.~\ref{fig:handbag}a as an example,
we thus have that the hard scattering, including radiative
corrections, cannot change the quark helicity.  In the collinear limit
helicity is equivalent to angular momentum along the collision axis,
so that the photon helicity must be conserved, too.  The leading Compton
amplitudes thus have a transverse initial photon, whereas the ones
with a longitudinal $\gamma^*$ are subleading in $1/Q$.  For meson
production one finds that the leading amplitude is the one where both
the $\gamma^*$ and the meson have helicity zero \cite{Collins:1996fb}.

These selection rules can be tested experimentally.  Consider the
electroproduction processes $ep\to ep\, \gamma$ or $ep\to ep\, M$ in
the targe rest frame.  The azimuthal angle $\varphi$ between the
planes spanned by the initial and scattered lepton and by the outgoing
proton and photon (meson) describes a rotation around the direction of
the intermediate $\gamma^*$.  Quantum mechanics tells us that
rotations are intimately connected with angular momentum, and it is
not surprising that the $\varphi$ distribution contains information
about the $\gamma^*$ helicity.  If the meson $M$ decays, as in
$\rho\to \pi^+\pi^-$, further information can be gained from its decay
angular distribution.

An application of these results is the $\sin\varphi$ dependence of the
single electron spin asymmetry in pion electroproduction measured by
HERMES \cite{Airapetian:2001iy}. It directly tells us that amplitudes
with a longitudinal and a transverse $\gamma^*$ have interfered.
Since the asymmetry is rather large we can infer that a subleading
amplitude (the one with a transverse photon) is not negligible
compared with the leading one for the process and $Q^2$ range in
question.  The angular analysis in deeply virtual Compton scattering
is more involved \cite{Diehl:1997bu}, but up to corrections a
$\sin\varphi$ behavior of the single electron spin asymmetry comes
from transverse and a $\sin(2\varphi)$ from longitudinal
$\gamma^*$. Both the HERMES and CLAS data \cite{Airapetian:2001yk}
shows little sign of a $\sin(2\varphi)$ variation and thus suggests
that the amplitudes with a longitudinal $\gamma^*$ are indeed
suppressed.

The arguments sketched above can be extended beyond the leading
approximation in $1/Q$.  For Compton scattering this has been worked
out, and one finds that at the level of $1/Q$ corrections one only has
longitudinal $\gamma^*$ polarization.  This is twice good news: for
one thing it means that $1/Q$ corrections (which need not be tiny at
moderate $Q^2$) do not affect the $\sin\varphi$ part of the single
spin asymmetry from which one wants to extract the leading amplitudes.
On the other hand, when data is sufficiently accurate to extract the
$\sin(2\varphi)$ component one will be able to study the first
subleading amplitudes for themselves.  This would be interesting
because they can still be expressed in terms of generalized parton
distributions at twist-three level, in a manner very similar to the
polarized structure function $g_2(x)$ in inclusive deep inelastic
scattering \cite{Anikin:2000em}.

\subsection{The target side}

Unless the transverse momentum transfer $(p'-p)_\perp$ is exactly
zero, the polarizations of the initial and final proton do not follow
selection rules like the photon and meson, they are indeed
unconstrained by the helicities on the photon side.  This is because a
momentum transfer of a few 100 MeV is small compared with $Q$, but not
compared with a typical hadronic scale governing the physics of the
hadronic matrix elements in our factorized diagrams.  The overall
helicity of the process need not be conserved since transfer of
orbital angular momentum will ensure angular momentum conservation.
This is another example of what we saw in the first section: by going
away from the strictly collinear kinematics of the usual parton model
quantities we get access to physics in the transverse direction, and a
particular aspect of this is the orbital angular momentum along the
longitudinal axis.

The wave function representation tells us for instance that the
generalized parton distribution $E$ (and thus also its first moment,
the Pauli form factor $F_2(t)$) would vanish if the proton did not
have configurations where the helicities of the partons do not add up
to the helicity of the proton.  This reflects from a quite different
angle the statement of Ji's sum rule \cite{Ji:1997ek} which connects
$E$ with the orbital angular momentum of quarks in the proton.

Since the hadron helicities are not fixed, a cross section in general
measures a particular combination of initial and final proton
helicities, and thus of the associated generalized parton
distributions.  In the large $Q^2$ limit, the beam charge asymmetry in
Compton scattering involves for instance the combination
\begin{equation}
\cos\varphi  \, 
\left[ F_1(t)\, H + \xi \Big(F_1(t) + F_2(t)\Big)\, \tilde{H} 
       - \frac{t}{4 M_p^2}\, F_2(t)\, E \right]
\label{comb-1}
\end{equation}
of Ji's distributions ($\tilde{E}$ drops out of this observable by an
accident of symmetry).  Notice that one cannot use the trick of
Rosenbluth separation here: to vary the factors weighting the
distributions in (\ref{comb-1}) one must vary $t$ or $\xi$, but the
distributions themselves depend on these variables.  On the other
hand, the full calculation \cite{Belitsky:2001ns} shows that with
longitudinal and transverse target polarization (but without measuring
recoil polarization) one obtains enough linear combinations of the
four distributions to separate them in principle.

\section{Generalized transversity distributions}

Quark transversity distributions, which in a helicity basis describe
quark helicity flip, are interesting objects for several
reasons \cite{Jaffe:2001dm}. Little is known about their size and
shape, and vigorous experimental effort will be devoted to improve
this situation in the coming years.  These distributions have of
course a generalization to non-forward
kinematics \cite{Hoodbhoy:1998vm}. Since helicity conservation does not
apply at nonzero $(p'-p)_\perp$ there are four distributions for each
quark flavor, corresponding to the four helicity combinations of the
initial and final proton \cite{Diehl:2001pm}. Time reversal invariance
does not eliminate any of these, but rather fixes their phase and
determines their behavior under $\xi\to -\xi$.  The counting of
generalized parton distributions for a given parton species and target
is thus different from the one for helicity amplitudes or for forward
distributions.

Unfortunately no process is presently known where generalized quark
transversity occurs, but the situation is better for the generalized
helicity flip distributions with gluons.  Note that these exist even
for spin $\frac{1}{2}$ targets, unlike their forward counterparts,
where the gluon helicity flip by two units must be matched on the
target side.  A remarkable property of these distributions is that
under evolution they do not mix with quarks (which cannot flip
helicity by two units).  Gluon transversity distributions can thus not
be generated by DGLAP-type gluon radiation off quarks and in this
sense probe ``intrinsic glue'' in the target.  Next to nothing is
known about them, but we can infer on their behavior at very small
momentum fractions from their evolution equations.  For this it is
enough to consider the forward case (with for instance a photon target
to satisfy helicity conservation) \cite{Ermolaev:1998jv}. In the double
logarithmic limit of large $\log Q^2\, \log(1/x)$ one obtains
\begin{equation}
g(x)        \sim x^{-1} f_1(x) , \qquad
\Delta g(x) \sim x^0    f_2(x), \qquad
\delta g(x) \sim x^1    f_3(x),
\label{hier}
\end{equation}
for the unpolarized, longitudinally polarized, and transversity
distributions, respectively, with the residual $x$ dependence in the
functions $f_{i}(x)$ controlled by $\alpha_s$.  Results of the double
logarithmic approximation should be used with care, but the hierarchy
in (\ref{hier}) does suggest that gluon transversity will fall and not
rise at small $x$.  The region of moderate momentum fractions may thus
be the best place to look for these distributions.

Gluon transversity occurs in deeply virtual Compton scattering, where
to leading order in $1/Q$ it provides the only contribution to the
photon double helicity flip amplitude.  It thus comes with a
characteristic signature in the $\varphi$ distribution, which provides
a handle to separate it from other distributions.  The beam charge
asymmetry to leading order in $1/Q$ for instance goes like
\begin{equation}
\cos3\varphi \, \frac{\alpha_s}{\pi}\,
\left[ F_2(t)\, H_T 
   - 2 \Big( F_1(t) + \frac{t}{4 M_p^2}\, F_2(t) \Big)\, \tilde{H}_T
   -  F_1(t)\, E_T \right] ,
\label{comb-2}
\end{equation}
where $H_T, \tilde{H}_T, E_T$ are three of the four gluon transversity
distributions.  Unfortunately this contribution has to compete with
terms suppressed by $M_p^2 /Q^2$ but without the factor
$\alpha_s/\pi$ \cite{Kivel:2001rw}. Whether gluon transversity can be
accessed in Compton scattering will thus depend on their size and on
the achievable values of $Q^2$.  Another process recently proposed is
$\gamma^* p \to \rho\, p$ \cite{Kivel:2001qw}. Here gluon transversity
appears in the amplitudes with double helicity flip from the photon to
the $\rho$, which are subleading in the $1/Q$ expansion.

\section{Deuteron target}

Some of the facilities that can study virtual Compton scattering and
meson production run with deuteron targets a part of their time, and
it is natural to ask what can be learned from the corresponding
generalized parton distributions.

The number of helicity combinations for a spin-one target is rather
large, but the algebra involved in the appropriate distributions is
not hopelessly tedious \cite{Berger:2001zb}. Concentrating on the
parton helicity conserving sector, a counting argument analogous to
the one we made for the proton tells us that for each parton species
there are 9 generalized distributions, 5 of them averaged over parton
helicities and 4 sensitive to parton polarization.

In the forward limit three of the quark distributions tend to the
well-known quark densities in the deuteron, which at leading order in
$\alpha_s$ are proportional to the structure functions $F_1(x)$,
$b_1(x)$, and $g_1(x)$.  Taking the lowest moment in $x$ we recover
the 3 vector and the 2 axial vector form factors for elastic deuteron
transitions from 5 of the generalized distributions.  The lowest
moments of the other 4 distributions turn out to be zero for different
symmetry reasons, one of them being directly related to the sum rule
for $b_1(x)$ found by Close and Kumano \cite{Close:1990zw}.

\begin{figure}[htbp]
\begin{center}
\leavevmode
\epsfxsize=0.98\textwidth
\epsfbox{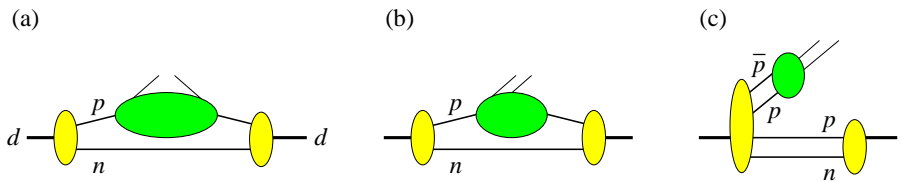}
\end{center}
\caption{\label{fig:deuteron} Different contributions to the
generalized parton distributions of the deuteron in a convolution
model.}
\end{figure}

The generalized parton distributions of a nuclear system are
interesting beyond their spin structure.  Consider the simple picture
of a deuteron as a weakly bound state of a proton and a neutron,
described by two-body wave functions with the appropriate momentum and
spin dependence.  As for usual parton densities, the parton
distribution in the deuteron is then a convolution of these nuclear
light-cone wave functions and the parton distributions in a single
nucleon, as shown in Fig.~\ref{fig:deuteron}a and \ref{fig:deuteron}b.
In this picture the momentum transfer of the process is entirely taken
by the emitting nucleon.  For nonzero $\xi$ the longitudinal momentum
fraction of this nucleon in the bound state before and after the
scattering is thus different, and the $\xi$ dependence of the
generalized parton distribution reflects rather directly the width of
the deuteron wave function in longitudinal momentum.  This dependence
is again accessible in correlation with the transverse distribution,
controlled by the variable $t$.  Since the deuteron is a weakly bound
system, its wave function will not allow too large values of
longitudinal transfer $\xi$.  For larger $\xi$ one will thus be
sensitive to quantum fluctuations of the deuteron that are more
complicated than a system of two almost free nucleons.

The story does not end here.  In analogy to the quark-antiquark region
of Fig.~\ref{fig:handbag}b, there is now the case where a
nucleon-antinucleon pair in the initial deuteron wave function
annihilates into a $q\bar{q}$ system which is then emitted
(Fig.~\ref{fig:deuteron}c).  The annihilation is described by a
generalized distribution amplitude, which can be studied in other hard
processes \cite{Muller:1994fv,Diehl:1998dk}. Intuition suggests that
four-nucleon configurations should be extremely rare in the deuteron,
but they are required by Lorentz invariance.  Whereas in the
description of the elastic form factors one can avoid their appearance
by choosing a frame with $\xi=0$, we now have nonzero $\xi$ imposed by
the process.  In this sense generalized parton distributions may
provide a handle to access truly relativistic effects in a nuclear
system.

\section{Conclusions}

Spin plays a variety of roles in the context of generalized parton
distributions.  In the hard processes processes where these quantities
appear, the helicity of the photon and meson provides a handle on the
reaction mechanism, which selects certain helicity transitions.
Measurement of the corresponding azimuthal distributions allows one to
separate different hard scattering contributions.  On the other hand,
the helicities in the hard subprocess do not fix those of the initial
and final hadron, and to separate their different combinations in the
parton distributions one needs target (or recoil) polarization.

A particular selection of photon helicities gives access to gluon
transversity distributions, provided power correction terms are
sufficiently small in the region under study.  Since they cannot be
generated from quarks via perturbative quark-gluon splitting, these
distributions constitute a rather unique probe of ``intrinsic glue'' in
the proton.  For spin $\frac{1}{2}$ targets they decouple in the
forward limit and are only visible in exclusive processes, when
helicity is not conserved.

The non-conservation of helicity in generalized parton distributions
is what makes these quantities sensitive to orbital angular momentum
in an essential way.  This is possible because one goes away from the
purely collinear kinematics describing the familiar twist-two
quantities of the parton model, and illustrates a characteristic
feature of nonforward exclusive processes: they simultaneously probe
the transverse and the longitudinal distribution of quarks and gluons
in a hadron state.  For nuclear systems we have seen that, in analogy,
one may be able to access the longitudinal and transverse distribution
of its nucleon constituents, and possibly more complicated
configurations.

Clearly, the dependence of generalized parton distributions on three
kinematical variables, and the number of distributions describing
different helicity combinations present a considerable complexity.  In
a sense this is the price to pay for the amount of physics information
encoded in these quantities.  It is however crucial to realize that
for many important aspects one need not fully disentangle this
complexity.  The interrelation of longitudinal and transverse structure
of partons in a nucleon, or of nucleons in a nucleus, can be studied
quantitatively from the distribution in the two external kinematical
variables $\xi$ and $t$, even if the deconvolution of the loop
integral over the variable $x$ is not performed.  In a similar way,
the very information of whether gluon transversity is small or large
for a typical size of longitudinal momentum set by $\xi$ would in
itself be valuable, given how little we know about the
non-perturbative origin of quark and gluon distributions in QCD bound
states.

\section*{Acknowledgements}

It is a pleasure to thank Bo-Qiang Ma and his colleagues for the
marvelous organization of this meeting, and Bernard~Pire for valuable
remarks on the manuscript.

\end{document}